# MULTI-VIEW TRANSFORMERS FOR AIRWAY-TO-LUNG RATIO INFERENCE ON CARDIAC CT SCANS: THE C4R STUDY

*Sneha N. Naik[1], Elsa D. Angelini[1,3], Eric A. Hoffman[4], Elizabeth C. Oelsner[1], R. Graham Barr[1], Benjamin M. Smith[1,2] and Andrew F. Laine[1]*

[1]Columbia University, USA. [2]McGill University, Canada. [3]Telecom Paris LTCI, Institut Polytechnique de Paris, France. [4]University of Iowa, USA.

## ABSTRACT

The ratio of airway tree lumen to lung size (ALR), assessed at full inspiration on high resolution full-lung computed tomography (CT), is a major risk factor for chronic obstructive pulmonary disease (COPD). There is growing interest to infer ALR from cardiac CT images, which are widely available in epidemiological cohorts, to investigate the relationship of ALR to severe COVID-19 and post-acute sequelae of SARS-CoV-2 infection (PASC). Previously, cardiac scans included approximately 2/3 of the total lung volume with 5-6x greater slice thickness than high-resolution (HR) full-lung (FL) CT. In this study, we present a novel attention-based Multi-view Swin Transformer to infer FL ALR values from segmented cardiac CT scans. For the supervised training we exploit paired full-lung and cardiac CTs acquired in the Multi-Ethnic Study of Atherosclerosis (MESA). Our network significantly outperforms a proxy direct ALR inference on segmented cardiac CT scans and achieves accuracy and reproducibility comparable with a scan-rescan reproducibility of the FL ALR ground-truth.

## 1. INTRODUCTION

Airway-to-lung ratio (ALR), defined as the ratio of mean airway lumen diameter measured at 19 standard anatomic locations divided by the cube-root of Total Lung Volume (TLV) measured at full inspiration (see Fig. 1 for details), is a computed tomography (CT)-based biomarker of airway "dysanapsis", which refers to a developmental mismatch in airway tree size relative to lung size [1]. CT-assessed ALR is associated with chronic obstructive pulmonary disease (COPD) risk later in life, explaining a similar proportion of airflow obstruction variance as tobacco smoking – the best known risk factor [2, 3]; is a risk factor for lung cancer, respiratory mortality, all-cause mortality in older adults [4] and is a key variable of interest in epidemiologic studies related to COVID-19 and the long-term health burdens of post-acute sequelae of SARS-CoV-2 infection (PASC) [5].

Gold-standard ALR measurement necessitates high resolution (HR, ~0.5mm voxel dimensions) CT-imaging capturing the full extent of the lung along with lung and airway tree segmentations with precise measures of TLV and airway lumen diameters at standard anatomic locations [3]. The Collaborative Cohort of Cohorts COVID-19 Study (C4R) has established COVID-19 and PASC risk in 49,979 participants from 14 retrospective cohorts [5]. The C4R CT study is missioned to measure ALR and TLV on 10 of these cohorts comprising n=12,459 HR full-lung but also n=13,752 cardiac CT images. Five of these cohorts only have pre-COVID cardiac CT scans, originally acquired for assessment of coronary artery calcium (CAC), including roughly 2/3 of the lung field of view with slice thickness of 2-3.5mm. Therefore, a method to estimate ALR from cardiac CT scans would roughly double the sample size for the C4R CT investigations, improving diversity, statistical power and generalizability of epidemiological analyses.

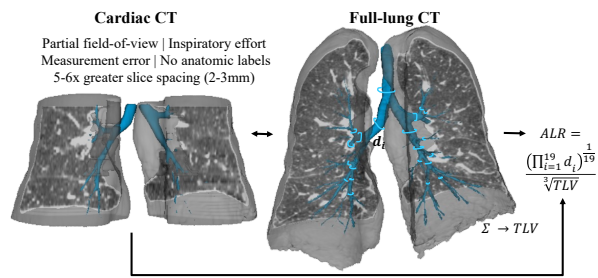

**Fig. 1:** 3-D rendering of cardiac and full-lung (FL) CT scans for a representative MESA Exam 5 participant with lung and airway segmentations. Ground-truth FL ALR is calculated as the ratio of the mean airway lumen diameter ($d_i$) across 19 specific anatomic locations to the cube root of the total lung volume (TLV) segmented on the full-lung scan.

Our previous work to quantify lung-structures on cardiac CT scans has successfully addressed pulmonary emphysema and its local phenotypes [6, 7] and TLV estimation from projections of the imaged lung and anatomical structures [8]. However, to the best of our knowledge, no prior literature exists to translate the measurement of airways widely studied on full-lung (FL) CT scans [9] to cardiac CT scans which presents the added challenge of segmenting and encoding fine structures from low resolution images. Attention-based vision-transformers and multiple-instance deep-learning (DL) methods have proven successful for learning patterns from large, high-resolution images with small dataset sizes across multiple medical image applications including CT, MRI and histopathology [10, 11]. We incorporate both building blocks in our proposed framework for ALR estimation. We trained a multi-view Swin-transformer [12] in two stages. First, we train Swin-transformer feature extractors to independently estimate ALR from axial, sagittal, and coronal projections of the cardiac CT airway and lung masks. Second, we aggregate features from all three views via a learned attention-weighting [11] to provide the final ALR estimation.

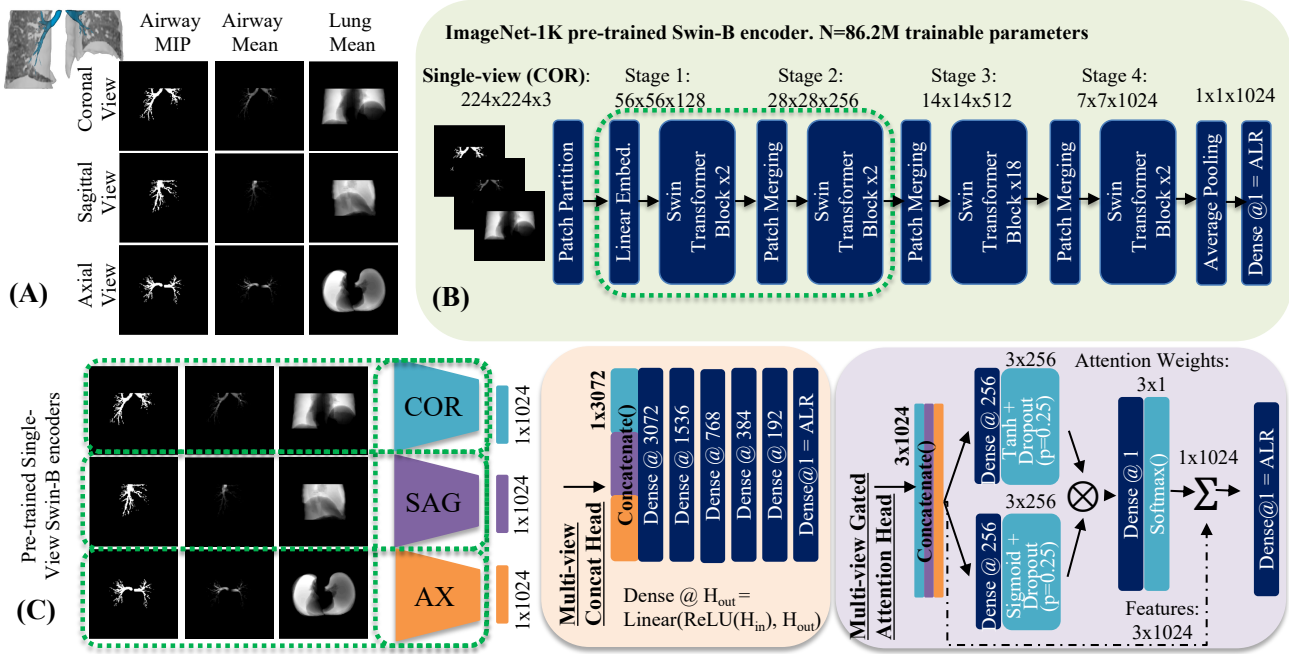

**Fig. 2 Schematic of Multi-View ALR Network:** (A) Generating single-view projections along Coronal (COR), Sagittal (SAG) and Axial (AX) views using the cardiac CT segmented airway and lung binary masks. (B) **Stage 1**: a network is trained per-view via fine-tuning a pretrained Swin-transformer (ImageNet-1K). Shallow network layers are frozen (green dashes). (C) **Stage 2**: three planar Swin-transformers are frozen (green dashes), and their latent features are either **concatenated** and passed to dense layers (orange box) or combined via learned **attention** weights (purple box) to infer the final multi-view ALR value.

We designed and validated our pipeline on thousands of paired FL and cardiac CT scans acquired as part of the Multi-Ethnic Study of Atherosclerosis (MESA) Lung study (a C4R cohort) at the same imaging session (Exam 5). We test the reproducibility of cardiac-inferred ALR on repeated cardiac scans acquired at the same imaging session (Exam 1) [13]. We demonstrate superior performance versus computing the cardiac imaged airway to imaged lung ratio. We compare our model's accuracy and reproducibility to published reports of ALR reproducibility on FL CT.

## 2. MATERIALS AND METHODS

### 2.1. Input datasets

The MESA Lung study is part of a population-based study of 6,814 participants recruited in 2000-2002 across six study centers in the USA [14]. Participants were aged 48-85, were 53% female, 39% white, and were free of cardiovascular disease at baseline. In 2000-2002 (Exam 1), two cardiac CT images were acquired for all consenting participants with ECG-gating, in-plane resolution of [0.547, 0.781]mm and slice spacing of [2.4-3.0] mm. In 2010-2012 (Exam 5), in addition to cardiac CT, participants underwent HR full-lung CT at full-inspiration with isotropic in-plane resolution between [0.467, 0.981]mm and slice spacing of 0.5mm [15].

In Exam 5, we identified n=1,885 participants with both cardiac and full-lung imaging, lung segmentations on both scans, airway segmentation and detailed airway lumen measures performed by VIDA Diagnostics Inc. (Coralville, IA, USA) on HR FL scans. The available Exam 5 participants were split 80:20 into a train and held-out test set, stratified by cardiac CT scanner model. The training set was further split 80:20 into a training and validation set, yielding n=1,206 training, n=302 validation and n=377 held-out participants for testing. The reproducibility set was n= 6,310 participants with repeated cardiac CTs in Exam 1 (without FL CT).

### 2.2. Cardiac CT Pre-processing

Cardiac airway segmentations were obtained via transfer learning from the paired HR FL scan in the training set. We generated a "simulated" cardiac CT and airway ground truth by cropping the FL CT images and corresponding airway segmentation masks to the intra-subject cardiac CT field-of-view and down-sampling along the superior-inferior axis as previously described [7]. A 2-D nnUNet segmentation model [16] was trained in a supervised manner using the simulated training dataset to segment visible airways with an equal weighted Dice plus Binary Cross-Entropy loss function [16], achieving a Dice Score Coefficient (DSC) of $0.87 \pm 0.03$ on the simulated validation set. The trained model was used to generate airway masks for the n=12.6k Exam 1 and n=1.8k Exam 5 cardiac CT images.

Each cardiac CT lung mask and airway mask was padded to the maximum field-of-view (FOV) observed in MESA Exam 5, then downsampled to a matrix size of 224x224x224 voxels with nearest neighbor interpolation. 2-D projections (array size 224x224 pixels) were generated from the 3-D volumes along axial, sagittal and coronal views consisting of (1) airway mask mean (airway silhouette), (2) airway maximum intensity projection (MIP), and (3) lung mask mean (lung silhouette). All three projections (Fig. 2A) were concatenated to generate a 224x224x3 input image per view. The ground-truth FL ALR values to infer were pre-processed by removing the training population mean value and scaling to unit variance. All available pairs of cardiac CT scans from Exam 1 (n=6,310) were preprocessed in the same way.

## 2.2. Model architecture and training

Our multi-view architecture was trained in two stages: In **Stage 1** (Fig. 2B), the shallow layers of an ImageNet pre-trained Swin Transformer [12] were frozen and partial fine-tuning was conducted using a single dense layer regression head to infer the ALR value. Models were trained per-view with AdamW optimizer, L2 regularization and a Cosine Annealing Warm Restarts learning-rate scheduler and mean squared error loss for 200 epochs with early stopping conditioned on validation loss. Images were augmented via left-right flipping with probability p=0.5 and rotation uniformly selected in the range of $\pm15°$.

We use the coronal single-view model (the one with best baseline initial performance) to jointly optimize the learning rate and L2-weight on the validation set by random search of 20 iterations. The optimal values (learning rate=$6.685\times10^{-5}$, L2 weight=$2.953\times10^{-4}$) are used for all subsequent experiments. For the coronal single-view model, we also perform an ablation study over data augmentation strategy and number of input projections.

In **Stage 2** (Fig. 2C), we tested view aggregation approaches via concatenation and gated attention. In the concatenation architecture we generate a single 3,072-dim feature vector per participant from the 3x1,024-dim per-view latent features from Stage 1. We train a custom dense neural network to infer the ALR value with five linear layers, each halving the feature size. The gated attention head (Eqn. 1, [11]) first computes attention weights ($a_k$) per view using two linear layers ($U, V$). A single participant-level feature vector ($z$) is generated from the weighted sum of the view-level features ($x_k$). Finally, a single linear layer ($W$) infers the ALR value. We optimized the architectures (depth, width, dropout) of both the dense concatenation and gated attention heads via a random sweep of 20 iterations to minimize validation mean squared error. Final architectures are shown in Fig. 2C. All other training parameters remained the same as in Stage 1.

$$a_k = \frac{exp\{w^T(\tanh(Vx_k^T)\odot sigm(Ux_k^T)\}}{\sum_{j=1}^{3} exp\{w^T(\tanh(Vx_j^T)\odot sigm(Ux_j^T)\}} \quad (1)$$
$$z = \sum_{k=1}^{3} a_k x_k, \quad ALR = Wz^T$$

## 2.2. Statistical Analyses

Model performance was quantified via the mean and standard deviation of the ALR residuals on the MESA Exam 5 test set, the $R^2$ correlation between predicted and ground truth ALR values and with Bland-Altman analysis. We measured the intra-class correlation (ICC) of predictions on repeated cardiac CT scans in MESA Exam 1, and quantified the variance explained ($R^2$ increment) for the standard spirometry-based measure of COPD physiology at MESA Exam 5 (the ratio of forced expired volume in 1 sec divided by forced vital capacity [FEV1/FVC]) [3]. We further benchmarked our deep model's performance against a simple "cardiac proxy" ALR, defined as the ratio of the mean lumen diameter across the largest 25 branches in the cardiac airway mask, to the cardiac lung mask volume. Airway branches were identified via skeletonization and bifurcation detection on the cardiac airway mask. Diameters were computed from the average cross-sectional area along the middle 2/3 of each branch (a similar approach is used to infer ground-truth ALR on FL CT scans [3, 17]).

## 3. RESULTS

The test-set mean, standard deviations and $R^2$ are presented in Table 1. In the coronal view, the three-channel model significantly outperformed estimates derived from airway maximum, airway silhouette or lung silhouette alone. Data augmentation via rotation and flipping led to an appreciable improvement in the three-channel coronal view model. Across the single-channel models, the coronal view outperformed sagittal or axial views alone.

The ALR estimated from both multi-view networks significantly outperformed the single-view networks (paired T-test, p-value $<10^{-17}$). View aggregation via gated attention yielded the best-performing DL model, with test-set residual standard deviation and $R^2$ reaching $0.00017\pm0.00181$ and $0.7413$ respectively. In addition, the model outperformed the correlation measures between cardiac- 'proxy' ALR and full-lung ALR in MESA Exam 5 by a significant margin. Bland Altman analysis (Fig. 3A) reported no significant fixed bias (p $>0.05$) and a small, but statistically significant proportional bias ($\beta$[95% CI] = -0.182 [-0.237, -0.127]), corresponding to underestimation of high FL ALR and overestimation of low

FL ALR values. Crucially, DL cardiac CT-inferred ALR explained a similar increment in FEV1/FVC variance explained (+8.7%, p<0.0001) after adjusting for risk factors (age, sex, race-ethnicity, primary/secondary smoke exposure, asthma) as the gold-standard full-lung ALR measure (+8.8%, p<0.0001) in the test-set, whereas the cardiac-"proxy" ALR measure, while still statistically-significantly, explained a 3.2% increment in FEV1/FVC variance. The prediction error of our best model is not significantly different between the smallest to largest quartiles of lung volume (unpaired T-test, p>0.05). Attention heatmaps (Fig. 3B) quantify the contributions of each view to the final ALR inference. For participants with high ALR, the coronal view features have higher weights whereas at low ALRs, sagittal and axial features have more weights. Inference on repeated cardiac scans in Exam 1 (Fig. 3C) showed excellent reproducibility of the ALR estimates (residual $R^2$=0.905, ICC=0.951), exceeding the reproducibility of the imaged lung volume ($R^2$ = 0.858), and comparing favorably with the FL 'gold-standard' ALR reproducibility (ICC=0.92 on n=96 with rescan interval 9-42 days [3]).

**Table 1:** Test-set performance of all Swin transformer models as single-view (Stage 1) or multi-view after features aggregation (Stage 2). Multi-input= Airway (AW) maximum, AW silhouette and lung silhouette. Augmentations= Rotation (R) ±15° and Left-Right flips (F). Mean/standard deviation (SD) of residuals and $R^2$ are reported versus 'gold-standard' full-lung ALR. Cardiac 'proxy' ALR metrics are also reported.

| | | Method | $R^2$ | Mean±SD |
|---|---|---|---|---|
| | | Cardiac 'proxy' | 0.5041 | 0.00182±0.00349 |
| Single-View | COR | AW max only, R&F | 0.4969 | -0.00008±0.00253 |
| | | AW silhouette only, R&F | 0.5189 | -0.00012±0.00247 |
| | | Lung silhouette only, R&F | 0.0415 | -0.00017±0.00354 |
| | | Multi-input, R only | 0.6662 | -0.00009±0.00207 |
| | | Multi-input, F only | 0.5123 | -0.00004±0.00249 |
| | | Multi-input, no R/F | 0.4633 | 0.000364±0.00261 |
| | | Multi-input, R&F | 0.7224 | -0.00021±0.00188 |
| | SAG | Multi-input, R&F | 0.6642 | -0.00019±0.00207 |
| | AX | Multi-input, R&F | 0.6759 | -0.00004±0.00203 |
| Multi-view | | Concatenation | 0.7337 | -0.00010±0.00184 |
| | | Gated Attention | **0.7413** | 0.00017±**0.00181** |

## 4. DISCUSSION

In this work, we have introduced a multi-view deep-learned network for inferring FL-based ALR values from segmented cardiac CT with only partial FOV and large slice thickness. Our proposed approach relies on single-view projections of segmented masks and single-view feature aggregation. Our results report high precision, reproducibility and yield reliable clinical ALR-measures with respect to the 'gold-standard' ALR on full-lung CT (the current strongest airway imaging biomarker for COPD-risk), paving the way for using our proposed cardiac-based ALR estimates in clinical studies. Our results outperform a direct estimate of a cardiac ALR using the airways segmented on cardiac CT, suggesting the ability of our projection views and deep network to adjust its inference with respect to the portion of airways and TLV included in the cardiac scan FOV. Varying inspiration levels and airway measurement errors on FL CT scans impose an upper bound on the accuracy of our ALR estimate. Future work will build on this model by exploring (1) training airway segmentation models directly on cardiac CT to limit protocol-related generalization error versus full-lung imaging and (2) adjusting our estimated ALR for participant demographics and BMI, scanner manufacturer, pulmonary emphysema, COPD status and severity, to improve predictive power and generalizability.

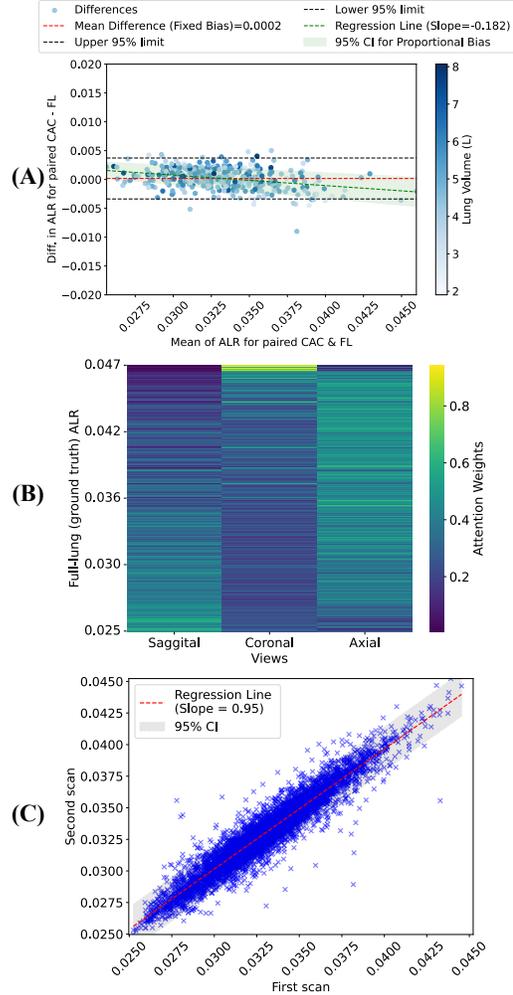

**Fig. 3:** Bland-Altman plot (A) for Exam 5 test-set predicted and ground truth ALR (n=377), from the best performing multi-view gated attention network, colored by TLV. Mean residual (red dashed) and 95% confidence interval (black dashed) are displayed. (B) Per view features learned attention weights on test-set examples. (C) Scatter plot of repeated ALR inferences for n=6,310 participants in MESA Exam 1.

## 12. COMPLIANCE WITH ETHICAL STANDARDS

Institutional review board approval was obtained for all study related activities. Written informed consent was obtained from all participants.

## 13. ACKNOWLEDGMENTS

This research was supported by the American Lung Association and by grants R01-HL121270, R01-HL077612, R01-HL093081, R01 HL-130506 from the National Heart, Lung, and Blood Institute (NHLBI) and OT2HL156812 from the NIH. MESA was supported by contracts 75N92020D00001, HHSN268201500003I, N01-HC-95159, 75N92020D00005, N01-HC-95160, 75N92020D00002, N01-HC-95161, 75N92020D00003, N01-HC-95162, 75N92020D00006, N01-HC-95163, 75N92020D00004, N01-HC-95164, 75N92020D00007, N01-HC-95165, N01-HC-95166, N01-HC-95167, N01-HC-95168 and N01-HC-95169 from the NHLBI, and by grants UL1-TR-000040, UL1-TR-001079, and UL1-TR-001420 from the National Center for Advancing Translational Sciences (NCATS). The authors thank the other investigators, the staff, and the participants of the MESA study for their valuable contributions. A full list of participating MESA investigators and institutions can be found at http://www.mesa-nhlbi.org. This publication was developed under the Science to Achieve Results (STAR) research assistance agreements, No. RD831697 (MESA Air) and RD-83830001 (MESA Air Next Stage), awarded by the U.S Environmental Protection Agency (EPA). It has not been formally reviewed by the EPA. The views expressed in this document are solely those of the authors and the EPA does not endorse any products or commercial services mentioned in this publication. Dr. Hoffman is a shareholder in VIDA Diagnostics, Inc.